\documentstyle[epsfig]{mn}
\title[The spatial correlation function of REFLEX clusters]
{The ROSAT-ESO Flux Limited X-ray (REFLEX) Galaxy Cluster Survey II: 
The Spatial Correlation Function \footnotemark[1]}
\author[C.A. Collins et al. ]
{C.A.~Collins$^1$, L.~Guzzo$^2$, H.~B\"{o}hringer$^3$,
P.~Schuecker$^3$, G.~Chincarini$^{2,4}$, R.~Cruddace$^5$ \newauthor
S.~De Grandi$^2$, H.T. MacGillivray$^6$, D.M.~Neumann$^7$,
S.~Schindler$^1$, P.~Shaver$^8$ \newauthor W.~Voges$^3$\\
$^1$Astrophysics Research Institute, Liverpool John Moores University,
Twelve Quays House, Egerton Wharf, Birkenhead, UK, CH41 1LD\\
$^2$Osservatorio Astronomico di Brera, via Bianchi 46, 22055 Merate
(LC), Italy\\ $^3$ Max-Planck-Institute f\"{u}r extraterrestrische
Physik, Gissenbachstra$\ss$e 1, 85740 Garching bei M\"{u}nchen, Germany\\
$^4$ Istituto dt Fisica Cosmica, CNR, via Bassini 15, 20133 Milano,
Italy\\ $^5$ E.O. Hulburt Center for Space Research, Naval Research
Laboratory, Code 7620, 4555 Overlook Ave., Washington, DC 29375, USA\\
$^6$ Royal Observatory, Blackford Hill, Edinburgh, EH9 3HJ, Scotland\\ 
$^7$ Service d'Astrophysique, CEA/Saclay, L'Orme des Merisiers Bat.
709, 91191 Gif-sur-Yvette, Cedex, France\\
$^8$ European Southern Observatory, Karl-Schwarzschildstrasse 2, 85748
Garching bei M\"{u}nchen, Germany}

\newcommand{\beq}{\begin{equation}}
\newcommand{\eeq}{\end{equation}}
\newcommand{\bfi}{\begin{figure}}
\newcommand{\efi}{\end{figure}}
\newcommand{\bit}{\begin{itemize}}
\newcommand{\eit}{\end{itemize}}

\newcommand{\myref}[1]{\noindent \hangindent=0.5in \hangafter=1 #1 \par}

\begin{document}

\maketitle
\begin{abstract}
We report the results of the spatial two-point correlation function
$\xi_{cc}(r)$ for the new X-ray galaxy cluster survey REFLEX, which
comprises of 452 X-ray selected
clusters (449 with redshifts) detected by the ROSAT satellite during the 
ROSAT All-Sky-Survey (RASS).  The REFLEX 
cluster sample is flux limited to $3\times10^{-12}$ erg s$^{-1}$ cm$^{-2}$ in the ROSAT energy band 
($0.1-2.4$ keV) and spans 3 decades in X-ray luminosity
($10^{42} - 10^{45} h^{-2}$ erg s$^{-1}$), containing galaxy
groups and rich clusters out to a redshift $z\leq0.3$. Covering a 
contiguous area of 4.24 sr REFLEX is the largest X-ray cluster sample
to date for which spatial clustering has been analysed.  Correlation
studies using clusters selected on the
basis of their X-ray emission are particularly interesting as they are
largely free from the projection biases inherent to optical studies.
For the entire flux-limited sample we find that the
correlation length (the scale at which the correlation amplitude
passes through unity) $r_0\simeq 20 h^{-1}$Mpc. For example, if a power-law 
fit is made to $\xi(r)$ over the range $4-40 h^{-1}$ Mpc then
$r_0=18.8\pm0.9$. An indication of the robustness of this result
comes from the high degree of isotropy seen in the clustering pattern
on scales close to the correlation length. 
On larger scales
$\xi_{cc}(r)$ deviates from a power-law, crossing zero at
$\simeq45h^{-1}$ Mpc. 
From an 
examination of 5 volume-limited cluster
sub-samples we find no significant trend of $r_0$ with limiting X-ray
luminosity. A comparison with recent model predictions for the
clustering properties of X-ray flux-limited samples, indicates that Cold
Dark Matter models with the matter density $\Omega_m=1$ fail to
produce sufficient clustering to account for the data, while  
$\Omega_m\simeq0.3$ models provide an excellent fit.
\end{abstract} 

\footnotetext[1]{Based on observations taken at The European Southern
Observatory, La Silla, Chile}

\section{INTRODUCTION}

Clusters of galaxies have been used for many years as tracers of the
large-scale mass distribution in the universe. As the largest
gravitationally bound objects their clustering statistics provide
important information on the hierarchical process of galaxy formation
enabling estimates to be made of the mass fluctuation amplitude and
the density paramter $\Omega$ (e.g. Mo et al. 1996). The early
statistical analyses relied on the visual cluster compilations
of Abell (1958) and Abell, Olowin \& Corwin (1989). From the redshift
surveys of richness-limited subsamples of the Abell catalogue (Bahcall
\& Soniera 1983, Klypin \& Kopylov 1983, Postman et al. 1992, Peacock
\& West 1992) it was established that the correlation function $\xi_{cc}(r)$
followed the form
\begin{equation}
\xi_{cc}(r)= \left ( \frac{r}{r_0} \right )^{-\gamma},
\end{equation}
on scales $\leq 100h^{-1}$ Mpc with $\gamma \simeq2$ and $r_0$
systematically 5 times higher than the value of $\simeq 5 h^{-1}$ Mpc
found for
galaxies (e.g. Davis \& Peebles 1983, Tucker et al. 1997), but with a 
strong dependency on the limiting richness of the cluster sample used 
(see Bahcall 1988). For
example, while the richness class $R\geq0$ samples give $r_0\simeq
20h^{-1}$Mpc, at the higher threshold $R\geq2$ the correlation length rises to
$r_0\simeq40h^{-1}$Mpc (Peacock \& West 1992). In principle these
results can be used to place constraints on theoretical models of
large-scale structure, however the cosmological
information they contain is questionable due to the likely existence
of inhomogeneities (Sutherland 1988, Sutherland \& Efstathiou 1991)
and line-of-sight projection effects (Lucey 1983, Dekel et al. 1990)
artificially enhancing the correlation amplitude of Abell-based
cluster samples. Strong evidence that these effects play a significant
role comes from comparing the amplitude of the correlation function
$\xi(\sigma,\pi)$ in the redshift direction $\pi$ of space with the
perpendicular direction $\sigma$. 
Both rich and poor Abell cluster samples regularly
fail this isotropy test, showing line-of-sight elongations in the
contours of $\xi(\sigma,\pi)$. These features are consistent with an
artificial
enhancement of the correlation function (Sutherland 1988, 
Efstathiou et al. 1992, Peacock \& West 1992) although physical 
interpretations have
also been suggested (Bahcall et al. 1986, Miller et al. 1999).

The advent of digitised cluster surveys saw a dramatic increase in the
homogeneity with which optical cluster samples could be compiled.
Results from both the Edinburgh/Durham Cluster Catalogue (Nichol et
al. 1992)
and the APM survey (Dalton et al.  1992, 1994, Croft et al. 1997)
demonstrated that for the equivalent Abell richness class $R\geq0$,
clusters found from automated detection algorithms have
$r_0\simeq15h^{-1}$Mpc
with significantly reduced anisotropies. Testing the results of the
richer clusters has proved more difficult due to the large search
volume required to find suitable numbers and the diminishing contrast
of distant clusters against the background of faint galaxies. For
example,
Croft et al. (1997) used 46 APM clusters with richnesses equivalent
to $R\geq2$ and found $r_0\simeq 20\pm5$.

In recent years attention has focused on cluster samples generated on
the basis of their X-ray emission. This method has enormous advantages
for the determination of $\xi_{cc}$ over the optically compiled cluster
catalogues described above:

\begin{itemize}

\item The X-ray emission from a cluster provides a direct physical
link with the presence of a large gravitational potential in
quasi-equilibrium (e.g. 
$L_x\propto M^{4/3}$). Thus the signature of X-ray emission provides
strong evidence that the apparent overdensities seen in the optical are
gravitationally bound structures.

\item The emissivity of Thermal Bremsstrahlung radiation is
proportional to the square of the electron number density,
whereas the optical richness estimates are simply proportional to the
galaxy density. 
Therefore, at fixed density, the contamination in cluster samples  
resulting from the 
projection of systems along the line-of-sight is intrinsically higher 
in richness-based optical samples compared to X-ray cluster catalogues. Furthermore the 
X-ray emission from clusters is concentrated towards the dense central cores which are
typically $\sim 250h^{-1}$ kpc in size -- significantly smaller
than the spatial extent of the galaxy concentration in clusters. Both
these effects substantially reduce the chance of projection effects 
which, as described above, are thought to plague Abell-based samples.

\item  The comparatively low internal background of the ROSAT Position
Sensitive Proportinal Counter (PSPC) and the relatively short exposure
times in the All-Sky Survey means that the X-ray fluxes from clusters at the 
flux limit of REFLEX are photon-noise limited as opposed to background
limited. This is in
contrast to purely optically selected samples which are forced to have a
minimum density contrast above a varying background of galaxies
before they can be detected.

\end{itemize}

The first attempts to measure $\xi_{cc}$ using X-ray clusters were
confined to
small samples: Lahav et al. (1989) detected significant clustering
with $r_0\sim21h^{-1}$ for $\gamma=1.8$ using an all-sky sample of 53
clusters
above a
flux $1.7\times10^{-11}$ erg s$^{-1}$ cm$^{-2}$ (2-10 keV). Nichol,
Briel \& Henry (1994) used ROSAT data for a complete sample of 67 X-ray 
bright Abell clusters finding a correlation length $r_0=16.1\pm3.4h^{-1}$ Mpc and
detecting no significant clustering anisotropy. A more extensive study
using data from the ROSAT satellite carried out by Romer et al. (1994) 
for a nearly complete flux-limited sample of 129 clusters above
$1\times10^{-12}$ erg s$^{-1}$ cm$^{-2}$ found
$r_0=12.9\pm2.2h^{-1}$Mpc, $\gamma=1.8\pm0.4$. This study also found no 
evidence of spatial anisotropy in the clustering pattern. More recently, there
have been two independent estimates of $\xi_{cc}$ from the 277 X-ray
brightest Abell cluster sample from the RASS (XBACS,
Ebeling et al. 1996). For this sample Abadi et
al. (1998) suggest $r_0=21.1^{+1.6}_{-2.3}h^{-1}$Mpc and $\gamma=1.9$
from a $\chi^2$ minimisation procedure using the binned correlation
data,
while Borgani et al. (1999) use a more reliable likelihood analysis
finding $r_0=26.0^{+4.1}_{-4.7}h^{-1}$ Mpc (here error bars are $2\sigma$). 
The anisotropy diagram for XBACS is
published by Miller et al. (2000) and shows strong Abell-type
elongations.

These X-ray results do not attempt to take account of the sky
coverage
of the parent X-ray survey in the correlation analysis. However, X-ray
cluster samples generated from the RASS (Tr\"{u}mper
1993, Voges et al. 1999) have the
advantage that the sky coverage is known from accurate information
on the X-ray flux limit pertaining to any part of the sky. 

The first
attempt to utilise the RASS sky coverage information is Moscardini et.
al. (2000a), who analyse the spatial distribution of the clusters in the 
RASS1 Bright Sample (De Grandi et al. 1999) using a very simple
version of the sky coverage based on the first processing of the
All-Sky Survey. This cluster catalogue is the forerunner to
REFLEX consisting of 130 clusters to a limit $3-4\times10^{-12}$ erg
s$^{-1}$ cm$^{-2}$ defined in the ROSAT hard energy band ($0.5-2.0$
keV) and covering an area covering 2.5 sr centred on the
Southern Galactic Cap. Moscardini et al. (2000a) find
$r_0=21.5^{+3.4}_{-4.4}h^{-1}$ and $\gamma=2.1^{+0.53}_{-0.56}$
($95.4\%$ errors) with a mild
dependence of $r_0$ on limiting flux and luminosity. The REFLEX survey
provides the opportunity to substantially improve on this result in a
number of important respects: (i) REFLEX
provides more than 3 times the number of X-ray clusters over a
contiguous area nearly twice as large. (ii) All RASS standard
analysis source detections are reanalysed using our own flux
determination method. (iii) Due account is taken of all exposure
variations, in contrast to the the RASS1 sample which is limited to
exposure times larger than 150 secs. (iv) The optical identification
is done in a homogeneous way based on the most comprehensive optical data base 
available for the southern sky. The power spectrum for REFLEX
is presented elsewhere (Schuecker et al. 2000), here we
concentrate on the correlation function.

The outline of the paper is as follows: In Section 2 we give a brief 
description of the REFLEX cluster survey. In Section 3 we discuss the 
algorithm used to esimate the the correlation
function and the results for both the entire REFLEX catalogue and
volume limited sub-samples are presented in Section 4. The
interpretation of these results in terms of structure formation models
is discussed in Section 5.  

\section{THE REFLEX Survey}

The REFLEX survey represents an objective flux-limited catalogue of
X-ray clusters in the southern hemisphere south of
declination +2.5 degs and excluding the region within $\pm20$ deg of
the Galactic Plane. A further $\simeq 324$ deg$^2$ of sky around the LMC and 
SMC is removed where X-ray detection is hampered by the high interstellar 
absorption and crowded star fields.
The remaining area covered by the survey is 13924 deg$^2$ or 4.24 sr, 
representing $\simeq34\%$ of the entire sky.

The primary X-ray data for REFLEX originates from the second
processing of the ROSAT
All-Sky-Survey (RASS2) using the Standard Analysis Software System (SASS) 
which is based on a maximum likelihood detection algorithm. Confirmed
RASS2 sources with a
likelihood parameter of at least 15 and count rate $\geq0.05$ cts
s$^{-1}$ in the $0.1-2.4$ keV energy band
have already been published in the RASS bright source catalogue (Voges
et al. 1999). For REFLEX we use the internal MPE source catalogue
totalling 54076 sources in the study area which allows the inclusion of sources with a
likelihood $\geq7$. Although some sources will be detected at a
significance $\leq 3 \sigma$ and not all are real, this lower
likelihood threshold ensures that the parent catalogue is as complete
as possible.

\begin{figure}
\vspace{-9.0cm}
\hbox{\hspace{0cm}\vspace{6.0cm}\psfig{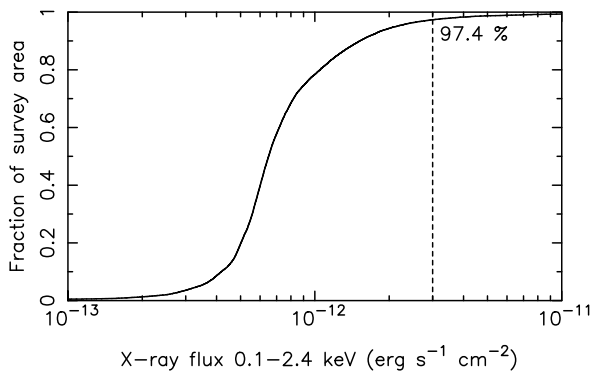}}
\caption{The sky coverage for REFLEX as a function of flux limit in
the ROSAT energy band. The curve is determined from the satellite
exposure map, the local hydrogen column density and the criterion that
at least 10 photons are detected. The dashed vertical line shows that
REFLEX reaches $3\times10^{-12}$ erg s$^{-1}$ cm$^{-2}$ for $97.4\%$
of the survey.}
\label{effective}
\end{figure}



It is well known from previous studies that the RASS analysis software
is optimised for point-like sources and therefore underestimates the flux 
of extended sources (e.g. Ebeling et al. 1996, De Grandi et al. 1997). Therefore
we have reanalysed all the source fluxes using a growth curve analysis
method to recover the total flux of extended sources with an internal
error of between $10-20\%$ (B\"{o}hringer et al. 2000a). Note: X-ray
count rates are measured in the hard band ($0.5-2.0$ keV) then
converted to unabsorbed fluxes in the ROSAT band ($0.1-2.4$ keV) and
the cluster X-ray luminosities are determined by an iterative
procedure using the luminosity-temperature relation of Markevitch
(1998) assuming $h=0.5$ 
(B\"{o}hringer et al. 2000c in preparation), with the values scaled by a
factor 0.25 to $h=1$ in this paper. 
Excluding double detections we have 4206 sources above a count rate
limit of 0.08 cts s$^{-1}$, which corresponds to a flux limit between
$1.6-2.0\times10^{-12}$ erg s$^{-1}$ cm$^{-2}$. 

The optical identification is based on finding 
galaxy overdensities in concentric rings around X-ray source positions using the UK 
Schmidt J-survey photographic plates digitised by COSMOS which reduces
the total number of cluster candidates to $\simeq500$ above a flux
limit $3\times10^{-12}$ erg s$^{-1}$ cm$^{-2}$ in the ROSAT energy band
$0.1-2.4$ keV. Details of the optical
identification process are given elsewhere 
(B\"{o}hringer et al. 2000b).  


To carry out further indentification and obtain redshifts,
multi-object ($5-20$ galaxies per cluster) and single-slit ($2-3$
galaxies per cluster) spectroscopy was carried out on $\simeq431$
targets as part 
of an ESO Key Programme (B\"{o}hringer et al. 1998, Guzzo et al.
1999). This results in 452 clusters above $3\times10^{-12}$ erg
s$^{-1}$ cm$^{-2}$ in the energy band $0.1-2.4$ keV, of which 449 have
secure redshifts either from our ESO programme or from the literature.
About $65\%$ of these clusters are in the Abell catalogue while most of the
others were previously unknown.

A comprehensive discussion of the contamination and
completeness statistics in REFLEX is 
given in B\"{o}hringer et al. (2000b) and results on
the comoving number density of clusters in the survey are presented in 
Schuecker et al. (2000). These indicate
a completeness well in excess of $\geq 90\%$ and a contamination by
non-cluster X-ray sources of less than $9\%$. We mention a few
results here to serve as an illustration of the quality of the catalogue: 
(i) From a search for X-ray emission around all ACO and ACO
supplimentary clusters only 1 cluster with an X-ray flux more 
than the flux limit is not found by the selection process. (ii) Clusters in
the luminosity range $0.08-2.5 \times10^{44}h^{-2}$ erg s$^{-1}$ have a 
constant comoving
number density of objects and  V/Vmax=$0.51\pm0.01$ at the flux
limit of the survey. This is consistent with the lack of evolution seen in
the X-ray luminosity function out to at least $z\simeq0.3$ reported by
other surveys ({\it e.g.} Burke et al. 1997, Ebeling et al. 1997). 
(iii) Approximately $81\%$ of the REFLEX
clusters are extended -- we searched the RASS2 database
separately for extended X-ray sources finding only a further 8 bona-fide
clusters and 5 candidate clusters, 3 of which show no obvious optical
counterpart and for which futher deep imaging is planned.








\section{Calculating the correlation function}

\subsection{Areal Coverage}

One complication with the RASS is that the sky coverage is not
homogeneous resulting in about $12\%$ of the REFLEX survey region
having an
exposure time less than half of the median exposure time ($\simeq 323$
s). This, coupled with the varying galactic hydrogen coloumn density, 
results in a
variation of the limiting flux of the RASS2 across the sky. Although the 
very low background for the ROSAT PSPC, especially in the
hard band ($0.5-2.0$ keV), allows the detection and characterisation of 
sources with
comparitavely low source source counts a minimum number is required
for a safe detection. Fig.~\ref{effective} shows the
resulting effective area of the REFLEX survey as
a function of flux with the additional criterion imposed of detecting
at least 10 photons in the hard band. The exposure times of the RASS2 in 
the REFLEX area 
are sufficient that at a flux limit of $3\times10^{-12}$ erg s$^{-1}$ 
cm$^{-2}$ at least 10 photons are detected for $97.4\%$ of the REFLEX
survey area and hence the number of clusters detected with low photon
counts is very small -- 3.8 clusters with less than 10 counts are expected
in the survey and only 1 is
detected. For a more conservative requirement of at least 30 photons
for each source the sky coverage falls to $78\%$ 
(see B\"{o}hringer et al. 2000b).

\subsection{Correlation Estimator}

In all computations we use the estimator

\begin{equation}
1+\xi_{cc}(r) = 4\frac{(DD) (RR)}{(DR)^2}, \label{Hamilton}
\end{equation}

where DD stands for the number of distinct pairs in the data, RR
stands for the number of distinct pairs in the random catalogue and DR
represents the number of cross pairs. The factor of 4 in this expression
accounts for the fact that while the number of distinct pairs in a large
catalogue of size $n$, say, is $\simeq n^2/2$, the number of cross-pairs
between two different catalogues each with $n$ entries is $\simeq n^2$
and these are all distinct. Since the correlation function is defined
in terms of the total number of pairs, the number of distinct $DD$ and
$RR$ pairs must each be multiplied by 2 to obtain the total numbers,
hence the factor of 4.
This estimator
has been shown by Hamilton (1993) to be the
most robust for datasets which may be sensitive to the chance
location
of strong clustering close to the sample boundary.

In principle the estimator in equation \ref{Hamilton} can be
generalised to
include an arbitrary weighting function. For calculating the correlation
function of galaxies from magnitude-limited samples, the variance in the
estimate of $\xi(r)$ on large scales is minimised if $w(r_i,\tau)$ is

\begin{equation}
w(r_i,\tau) = \frac{1}{1+4 \pi n_D J_3(\tau) \phi(r_i)},
\end{equation}

where $r_i$ is the distance of an object from the origin, $\tau$ is
the distance separating two objects, $\phi(r)$ is the survey selection 
function, $n_D$ the mean space density of objects and 
$J_3(\tau)=\int_{0}^{\tau} dr r^2 \xi(r)$ (see Saunders,
Rowan-Robinson \&
Lawrence 1992, Fisher et al. 1994, Guzzo et al. 2000). The
physical
interpretation of the term $4 \pi n_D J_3(\tau)$ in the weighting
scheme is that it represents something like `the mean number of
objects per
clump'. For galaxies this number is large on small scales giving
equal volume weighting to the pairs ($w\propto1/\phi(r)$). 
By
contrast, for
clusters the weighting term $4 \pi n_D J_3(\tau)$  is always
small compared to unity on scales of interest and consequently we
assign equal weight to all pairs in the calculation of the cluster
correlation function.

We calculate spatial separations using the formula for comoving
coordinate distance ($r_1$);

\begin{equation}
r_1 = \frac{\rm{c}}{\rm{H}_0}
\left[\frac{(q_0z)+((1-q_0)(1-((2q_0z)+1)^{1/2}))}{(1+z)q^{2}_0}\right],
\label{distance}
\end{equation}

adopting the cosmology $\rm{H}_0=100h^{-1}$ Mpc, $\Omega_m=1.0$ \&
$\Omega_{\Lambda}=0.0$, along with the
cosine rule to determine angular separations.

\subsection{Random catalogues}

The random catalogues are constructed over the REFLEX survey area
using a
Monte-Carlo technique which incorporates
knowledge of the flux limit in cells of size $\simeq 1$ square degree.
To begin with it is assumed that the observed number count
distribution of X-ray
clusters, LogN-LogS, is well fitted by a simple lower-law:

\begin{equation}
N(> S) = A S^{-\alpha}.
\end{equation}

For the purposes here we adopt
the value $\alpha=1.35$, consistent with the REFLEX number counts (see
B\"{o}hringer et al. 2000b) and those of RASS1, the precursor survey of REFLEX 
(De Grandi et al. 1999). Small changes to the value of $\alpha$ does not 
alter the outcome of the results. If we assume that the accumulative
distribution $P$ defined as

\begin{equation}
\frac{N(<S)}{N_{total}(>S_{lim})}=P
\end{equation}

is uniformly distributed in the range $0\rightarrow1$, then the
distribution of cluster X-ray fluxes $S$ selected at random above
$S_{lim}$ is given by

\begin{equation}
S = S_{lim}(1 - P)^{-\frac{1}{\alpha}}. \label{simulate}
\end{equation}

\begin{figure}
\vspace{-9.0cm}
\hbox{\hspace{0cm}\vspace{6.0cm}\psfig{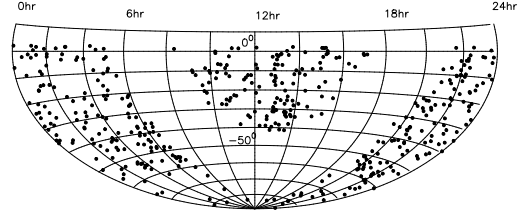}}
\caption{Aitoff plot of 449 REFLEX Clusters with redshift.}
\label{dataplot}
\end{figure}

\begin{figure}
\vspace{-9.0cm}
\hbox{\hspace{0cm}\vspace{6.0cm}\psfig{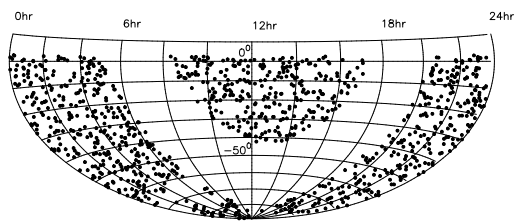}}
\caption{Aitoff plot of 1000 random points generated with the REFLEX
sensitivity map (flux limit $3\times10^{-12}$ erg s$^{-1}$ cm$^{-2}$
with a minimum of 10 photon counts) and mask.}
\label{ranplot}
\end{figure}

\begin{figure}
\resizebox{8.0cm}{!}{\includegraphics{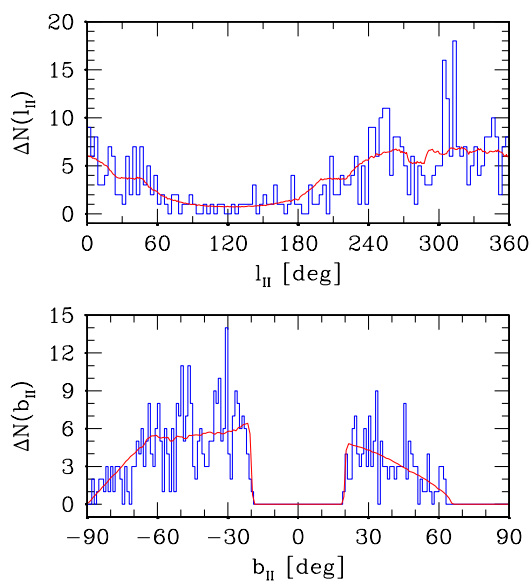}}
\caption{Histogram of REFLEX clusters as a function of galactic
longitude and latitude. The solid line is the prediction from a random
distribution of points convolved with the REFLEX sensitivity map (flux
limit $3\times10^{-12}$ erg s$^{-1}$ cm$^{-2}$ with a minimum of 10
photon counts) and mask.}
\label{lb}
\end{figure}

We select a cluster at random within the allowed borders of the REFLEX
survey and then use eqn.~\ref{simulate} to assign it a flux. We then
test whether the cluster falls above or below the flux limit for that
region of the REFLEX survey based on the local values of exposure time
and the interstellar hydrogen 
column density (Dickey \& Lockman 1990, Stark et al. 1992). We set
$S_{lim}=3\times10^{-12}$ erg s$^{-1}$ cm$^{-2}$, which is the
cut-off flux limit for the entire survey. An example of a random catalogue
with 1000 points generated in this way is shown in Fig.~\ref{ranplot}. To
demonstrate reliability of the random catalogues the histogram in 
Fig. ~\ref{lb} shows
the number of REFLEX clusters as a function of Galactic longitude and
latitude compared to that of a random catalogue generated using the
REFLEX survey sensitivity map. The random catalogues used in the
determination of $\xi_{cc}$ contain typically 100,000 points.

\begin{figure}
\vspace{-9.0cm}
\hbox{\hspace{0cm}\vspace{6.0cm}\psfig{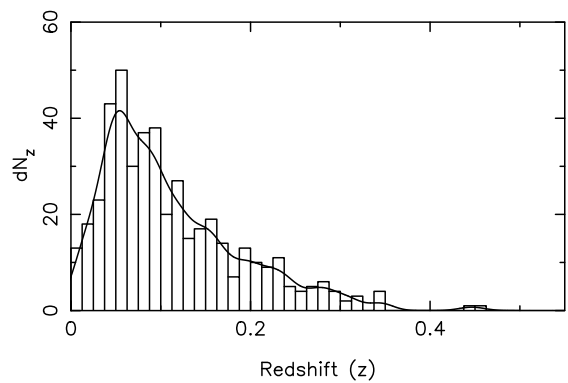}}
\caption{Histogram of the 449 REFLEX clusters with redshifts. The
solid line is the estimated density distribution using a Gaussian
kernel of width 5600 km s$^{-1}$.}
\label{smooth}
\end{figure}

\begin{figure}
\vspace{-9.0cm}
\hbox{\hspace{0cm}\vspace{6.0cm}\psfig{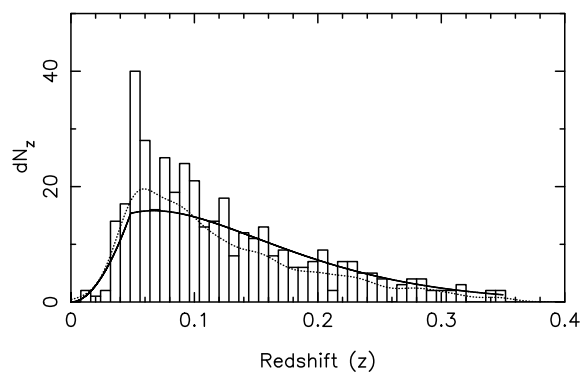}}
\caption{Histogram of the 399 REFLEX clusters with $L_x \geq 3 \times 
10^{44}$ erg s$^{-1}$. The dashed curve is the estimated redshift distribution of this sample using 
a Gaussian kernel of width 5600 km s$^{-1}$, as for Fig.~\ref{smooth}.
The solid curve is the redshift distribution estimated by 
integrating the X-ray cluster luminosity function.} 
\label{ranzlf}
\end{figure}




We use two methods to assign each random point a redshift. (i) Redshifts 
are drawn from the distribution of REFLEX clusters smoothed with a Gaussian 
kernel. This method allows the redshift selection function to be estimated
without prior knowledge of the underlying density distribution of
clusters and is used in almost all previous determinations of the cluster
correlation function. For REFLEX we use a Gaussian of width 5600
km s$^{-1}$ -- the optimum value depends on the space density of 
clusters and is constrained by the need to follow the redshift distribution 
accurately enough while not removing large-scale clustering. The exact 
figure used is generally not critical, with values in
the literature ranging between $4000-8000$ km s$^{-1}$. 
(ii) The second method, which we apply to luminosity limited samples,
uses the X-ray cluster luminosity function to generate the expected
number of clusters at each redshift. Assuming a Schechter function of 
the form

\begin{equation}
n(L)dL = A exp(L/L_\star) (L/L_\star)^{-\alpha} dL,
\label{schechter}
\end{equation}

where $n(L)$ is the number density of clusters per luminosity interval, 
then for particular values of $\alpha$ and $L_\star$, we can integrate $n(L) dL$ 
above $L_{lim}$ to determine the number density of
clusters at each redshift $\eta(z)$. The value of $L_{lim}$ at each redshift is
found from the flux limit (fixed at $3\times10^{-12}$erg
s$^{-1}$ cm$^{-2}$). The expected number of clusters in each redshift
interval $dz$ is then simply $\eta(z)*dV(z)$, where $dV(z)$ is the comoving
volume element at redshift $z$. Each random point generated in the
area of the survey is thus assigned a
random redshift weighted by the expected number of clusters based
on eqn.~\ref{schechter}. We have used the values of
$\alpha=1.61$, $L_\star=6.04\times10^{44}$ (erg s$^{-1}$), 
$A=3.04\times10^{-8}$ ($10^{44}$ erg s$^{-1})^{-1}$ and
H$_0=50$ km s$^{-1}$ Mpc$^{-1}$, which are appropriate to the REFLEX
sample (B\"{o}hringer et al. 2000c, in preparation), although using
our previous luminosity function parameters from De Grandi et al.
(1999) gives
identical results. The redshift distribution for the REFLEX
sample along with the smoothed version using method (i) is shown 
in Fig.~\ref{smooth}, while Fig.~\ref{ranzlf} shows a comparison
between the redshift distribution of both methods for the luminosity 
subsample of 399 clusters limited to $L_x\geq 3\times10^{44}$erg s$^{-1}$. 
We prefer
method (ii) for the luminosity sub-samples as it makes no prior
assumptions regarding the scale of the clustering and avoids the need
to smooth the data, however in practice
we found no significant difference in the correlation functions
resulting from the two methods.

\subsection{Maximum Likelihood Determination of $r_0$ and $\gamma$}

In calculating the best-fit power-law for the correlation function from
samples of
$\sim 100$ clusters there has
traditionally been one of two methods adopted. The first is to
calculate errors for $\xi_{cc}$ based on estimates from pair counts
binned
into $\simeq$ 10 coarse intervals, which are usually spaced
logarithmically
out to $\simeq 100h^{-1}$ Mpc ({\it e.g.} Bahcall et al. 1983, Dalton et al. 1992,
Nichol et al. 1992). On the grounds that each bin contains a large
numbers of pairs, the best-fit
power-law is estimated using the $\chi^2$ statistic. The danger
with such an approach is that the resulting estimates of parameters
describing the power-law ($r_o, \gamma$) will
then depend on the precise details of the binning. In order to
overcome this
limitation we adopt the second of the two methods referred to above and
maximise the likelihood $L$ that the model correlation function
produces the
measured number of cluster pairs at a given separation
(Croft et al. 1997, Borgani et al. 1999, Moscardini et al. 2000a). The likelihood
estimate is based on Poisson probabilities, such that

\begin{equation}
L=\prod_{i=1}^{N}e^{-\mu}  \mu^{\nu}/\nu!,
\end{equation}

where $\nu$ is the observed number of cluster-cluster
pairs in a small interval $dr$ and $\mu$ is the expected number in the
same interval calculated using Hamilton's estimator
(eqn.~\ref{Hamilton}).
As long as the number of random points is kept large enough to avoid a
zero in the denominator of eqn.~\ref{Hamilton}, $dr$ can be made
arbitrarily small, so as to ensure the final results are independent
of the bin size. In practice we used $\sim 7000$ bins between $5-100$
Mpc,
which resulted in either a 0 or 1 cluster-cluster pair in almost all bins.

The associated errors on the correlation function are usually calculated
from `Poisson' statistics. In the case of large bin intervals errors
are computed from the formula

\begin{equation}
\delta\xi_{cc}(r)=\frac{(1+\xi_{cc}(r))}{\sqrt{N_{cc}}},
\label{poisson}
\end{equation}

where $N_{cc}$ is the number of distinct cluster pairs in the bin
centred at separation $r$. In the case of a maximum likelihood
determination, such as that used here, confidence levels can be
defined as $S(r_{best},\gamma_{best})-S(r_{0},\gamma)$, where $S$ is
the usual $S=-2ln L$, assuming that $\Delta S$ is distributed like
$\chi^2$.

Both these methods are likely to produce underestimates of the true
dispersion as the use of Poisson statistics assumes that the pair
counts are independent of each other, which is clearly not the case.
An estimate of the true dispersion in the correlation function, which
tries to account for cosmic variance, can be made either by
applying a bootstrap resampling of the real data
({\it e.g.} Ling, Frenk \& Barrow 1986; Mo, Jing \& B\"{o}rner 1992)
or by carrying out numerical simulations based on plausible
cosmological models (Croft \& Efstathiou 1994, Croft et. al.
1997). Both methods produce similar results indicating that the real
errors are probably $1-2$ times larger than the Poisson-based estimates.
We confirmed this result for our sample by generating mock catalogues 
from bootstrap resampling of the data and calculating the
best-fit $r_0$ and $\gamma$ values using the likelihood method
described above. The ratio of the error for $r_0$ from the variance
between the bootstrap samples and the Poisson error is between
$\simeq1.5-2.0$ for all 
luminosity subsamples. 
Unless stated otherwise, in the results which follow we quote the $1
\sigma$ likelihood errors on the values of
$r_0$ and $\gamma$.


\section{Results}

\begin{figure}
\vspace{-9.0cm}
\hbox{\hspace{0cm}\vspace{6.0cm}\psfig{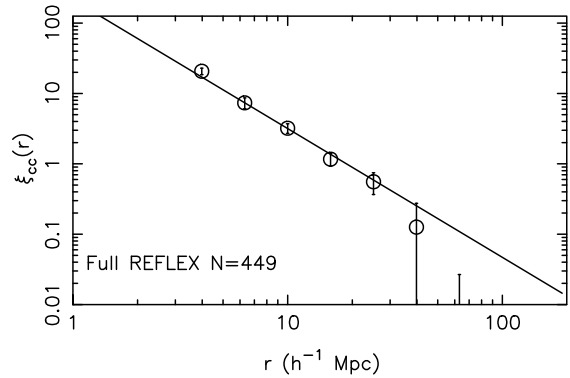}}
\caption{The correlation function for the REFLEX survey of 449 clusters. 
The error bars on each point are derived from bootstrap statistics.
The solid line shows the result of the likelihood analysis
($r_0=18.8$, $\gamma=1.83$) fitting a power law over the range
$4-40h^{-1}$ Mpc.}
\label{allkpcorrel}
\end{figure}

\begin{figure}
\vspace{-9.0cm}
\hbox{\hspace{0cm}\vspace{6.0cm}\psfig{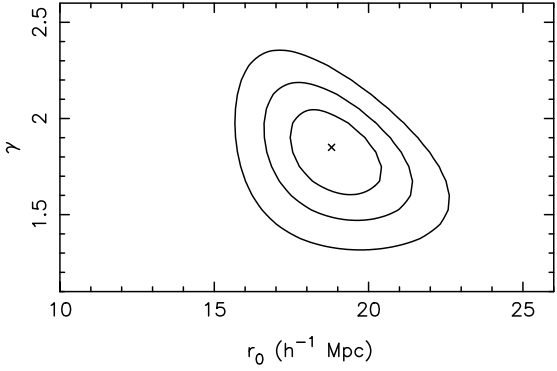}}
\caption{The probability contours ($1\sigma, 2\sigma, 3\sigma$) for the best 
fit $r_0$ and $\gamma$ from the likelihood 
analysis of the REFLEX survey over the range $4-40h^{-1}$ Mpc.}
\label{likeall}
\end{figure}

\begin{figure}
\vspace{-9.0cm}
\hbox{\hspace{0cm}\vspace{6.0cm}\psfig{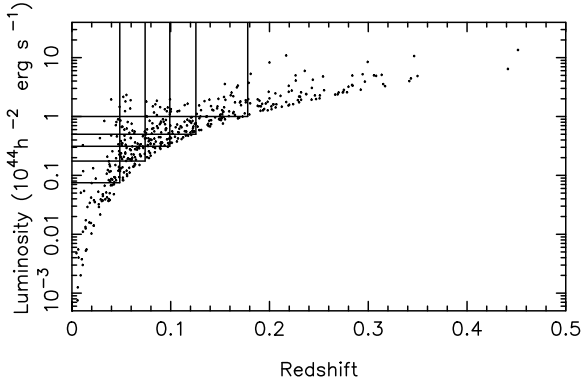}}
\caption{The X-ray luminosity ($0.1-2.4$ keV) vs redshift for the 
REFLEX sample of 449 clusters. Also shown are the 5 volume complete
sub-samples used to examine the variation of the correlation length
with limiting X-ray luminosity ($L_{lim}$.}
\label{lxvsz}
\end{figure}

The $\xi_{cc}(r)$ for the REFLEX survey of 449 clusters is shown in
Fig.~\ref{allkpcorrel}. A fit was made to the correlation function 
assuming a single power law over the range $4-40h^{-1}$ Mpc using the 
likelihood analysis described in
Section 3.4. Fig.~\ref{likeall} shows the corresponding joint
constraints resulting from this analysis. The best-fit value
for the power-law parameters are $r_0=18.8\pm0.9$ and
$\gamma=1.83^{+0.15}_{-0.08}$. If points are included 
on larger scales then the slope steepens, e.g. fitting between 
$4-100h^{-1}$ Mpc gives $\gamma\simeq2.35$ and $r_0\simeq16.25$. The
inability of a single power law to adequately describe the correlation
function is
further reflected in the zero crossing of $\xi_{cc}$ at
$45h^{-1}$ Mpc (see Section 7.1). 
 
\begin{figure}
\vspace{-9.0cm}
\hbox{\hspace{0cm}\vspace{6.0cm}\psfig{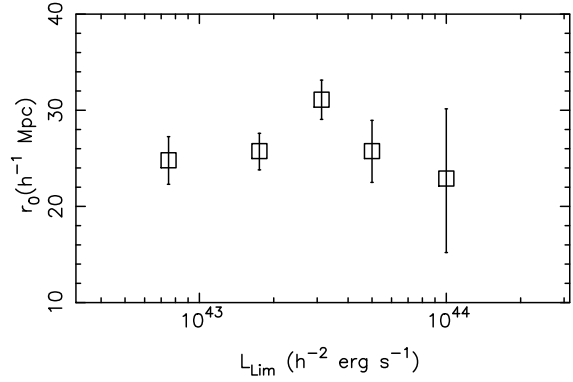}}
\caption{The correlation function amplitude $r_0$ plotted against
limiting X-ray luminosity defined in the ROSAT energy band ($0.1-2.4$
keV).
The errors correspond to $1 \sigma$ from the likelihood analysis.}
\label{rovslx}
\end{figure}
\begin{figure}
\vspace{-9.0cm}
\hbox{\hspace{0cm}\vspace{6.0cm}\psfig{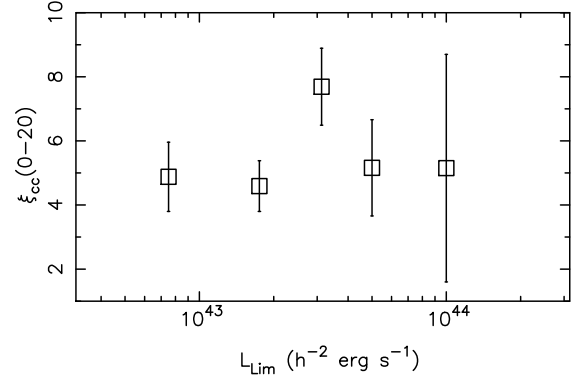}}
\caption{The correlation function amplitude in the range $0-20h^{-1}$ Mpc 
plotted against limiting X-ray luminosity defined in the ROSAT energy band ($0.1-2.4$
keV). The error bars are $1 \sigma$ based on eqn.~\ref{poisson}.}
\label{xi020vslx}
\end{figure}

In order to investigate the dependency of $r_0$ with X-ray luminosity 
we also calculated $\xi_{cc}(r)$ for 5
volume-limited X-ray sub-samples with luminosity thresholds
$0.08,0.18,0.3,0.5,1.0$ in units of $10^{44}h^{-2}$ erg s$^{-1}$.
Fig.~\ref{lxvsz} shows the distribution of luminosity with redshift
for the REFLEX sample along with the regions corresponding to the 
5 subsamples. Due to the significant covariance between $r_0$ and
$\gamma$ this investigation has been carried out with $\gamma$ fixed
at 2.0. The correlation results for the sub-samples
are presented in Table~\ref{rogamma} and Fig.~\ref{rovslx}. These
indicate no significant positive trend of $r_0$ vs $L_{lim}$, 
with the highest measured $r_0$ occuring at intermediate luminosities 
($L_{lim}\geq0.3\times10^{44}h^{-2}$ Mpc). Beyond this point the
statistical errors increase rapidly. To test the reliability of the
parameter $r_0$ as an
indicator of how clustering changes with X-ray luminosity, we calculated the
average correlation function amplitude for the 5 volume-limited sub-samples over  
the range of separations $0-20h^{-1}$ Mpc. The result, shown in
Fig.~\ref{xi020vslx}, is in good agreement with the trend of $r_0$ vs
$L_{lim}$.

\begin{table}
\caption[]{Estimates of r$_0$ as a function of limiting X-ray
luminosity for volume limited cluster subsamples. Values of r$_0$ are
calculated at $\gamma=2.0$}
\begin{tabular}{l|c|r|}\hline
$L_{lim} (10^{44}h^{-2}$) erg s $^{-1}$  &  Number & $r_0$  \\\hline
1.0 & 67 & $22.9^{+7.3}_{-7.7}$  \\
0.5 & 101 & $25.8^{+3.2}_{-3.3}$ \\
0.3 & 108 & $31.1^{+2.0}_{-2.1}$ \\
0.18 & 84 & $25.8^{+1.9}_{-2.0}$ \\
0.08 & 39 & $24.8^{+2.5}_{-2.5}$ \\\hline
\end{tabular}
\label{rogamma}   
\end{table}

\begin{figure}
\vspace{-9.0cm}
\hbox{\hspace{0cm}\vspace{6.0cm}\psfig{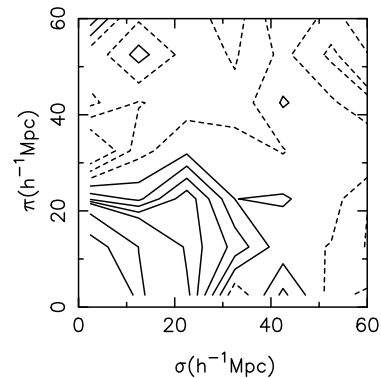}}
\caption{Contours of constant $\xi(\sigma,\pi)$ for the REFLEX survey.
The contour values used are 4.0, 2.0, 1.0 and then 0.8 to -0.4 in steps of 0.2}
\label{pisig}
\end{figure}


We have examined the isotropy of the clustering signal for the REFLEX
survey by plotting contours of $\xi(\sigma,\pi)$, where
$\pi=|r_1-r_2|$ is the line-of-sight separation, with $r_{1,2}$
determined from eqn.~\ref{distance}, and
$\sigma=(s^2-\pi^2)^{1/2}$ is the perpendicular component of the
cluster separation $s$. As discussed in the introduction, 
elongations of the contours in the
redshift direction compared to the perpendicular direction for scales 
$\simeq20h^{-1}$) Mpc are a
feature of some optical cluster catalogues -- typically with a ratio
$\simeq4:1$ for redshift samples based on the Abell
catalogue (e.g. Postman et al. 1992).
Fig.~\ref{pisig} represents the corresponding plot for the REFLEX
clusters and indicates that unlike optical surveys, the
$\xi(\sigma,\pi)$ contours are close to being completely concentric on
scales close to the correlation length.






\section{Discussion}

The determination
$\xi_{cc}$ from the REFLEX survey can be compared with similar
determinations for other X-ray cluster samples. Our results of
$r_0=18.8$ and little dependency of $r_0$ on X-ray luminosity 
are broadly consistent with the results of XBACS (Borgani et al. 1999)
and RASS1 (Moscardini et al. 2000a). The result presented by Romer et
al. (1994), hereafter R94, for a sample of 128
clusters above $1.0\times10^{-12}$ erg s$^{-1}$ cm$^{-2}$ in a 3100 deg$^2$ 
area centered on the SGP, suggests a correlation
length $r_0=13-15 h^{-1}$ Mpc, smaller than any of the other determinations from
X-ray samples. In
addition to the fainter flux limit, the R94 study differs from REFLEX in 2 further  
ways which in principle could affect the result: (i) the cluster 
sample was based on a reduction of the
all-sky-survey using the ROSAT Standard Analysis Software, which has 
subsequently been revised (ii) the correlation analysis performed by
R94 did not include the sample sky coverage
corresponding to the SGP region under study. We have investigated the 
origin of a possible systematic difference between R94 and REFLEX by
repeating our correlation analysis on the 109 REFLEX clusters lying 
within the R94 SGP area of sky, defined by the 
boundaries
$22\rm{hr}\leq\rm{RA}\leq3 \rm{hr}$, $-50^\circ \leq \rm{dec} \leq
2^\circ$,$ |b| \geq 40^\circ$. The resulting power-law fit to the
correlation function out to $\leq100 h^{-1}$ Mpc gives
$r_0=12.9^{+1.9}_{-1.9}$ $\gamma=2.0^{+0.4}_{-0.4}$, smaller than the
REFLEX amplitude of $18.8\pm0.9$ and very close to
the original SGP result of $r_0=12.9\pm2.2\, h^{-1}$
Mpc, $\gamma=1.8\pm0.4$ found by R94 fitting over the
same range. This suggests 
that the difference between the REFLEX and the SGP result is most likely 
due to the superior statistical sampling of REFLEX which represents a 4-fold 
increase in survey area over the SGP region while probing to a similar
redshift.

Miller et al. (2000) analyse the $\xi(\sigma,\pi)$
diagram for a number of X-ray cluster samples. In their analysis the
XBACS clusters show very strong elongations around
$\xi(\sigma,\pi)\simeq1$ in the redshift direction and a similar 
anisotropy is present in other X-ray
confirmed Abell cluster samples. The RASS1 sample shows a much weaker
anisotropy over the same scale.
On the basis of this Miller et al. (2000)
argue that clustering anisotropy is a ubiquitous feature of
X-ray cluster samples which demonstrates that the anisotropies are real.
However, the absence of any significant anisotropy
in the contours of
$\xi(\sigma,\pi)$ for the REFLEX survey shown in fig.~\ref{pisig}
indicates that this is not the case.
This is the strongest indication yet that
elongations seen in other
catalogues are spurious and justifies the claim, first pointed out by
R94, that X-ray selected surveys do not suffer from
significant projection biases. 
As with
optical studies based on the Abell catalogue, the presence of strong 
elongations close to the scale of $r_0$ in the XBACS bring the 
accuracy of the clustering signal derived from this sample into question.


\subsection{Comparison with Cosmological Models}

\begin{figure}
\vspace{-9.0cm}
\hbox{\hspace{0cm}\vspace{6.0cm}\psfig{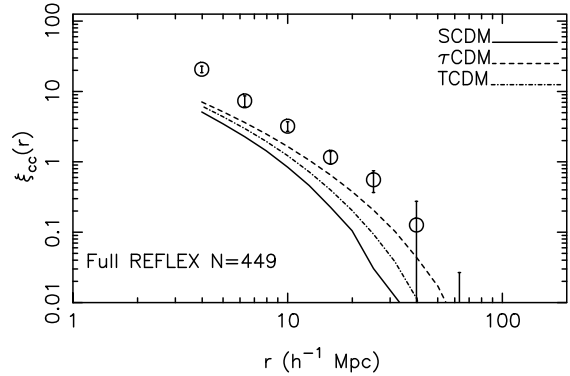}}
\caption{The REFLEX correlation function compared with a range
of Cold Dark Matter models for which $\Omega_m=1$ and
$\Omega_{\Lambda}=0$, using the same flux limit and sky coverage as
REFLEX -- taken from Moscardini et al. (2000b). Error
bars in this figure are $1 \sigma$ and
based on
bootstrap resampling.}
\label{cdm1}
\end{figure}

\begin{figure}
\vspace{-9.0cm}
\hbox{\hspace{0cm}\vspace{6.0cm}\psfig{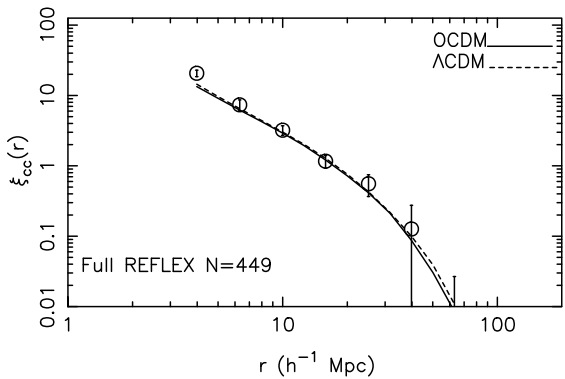}}
\caption{The REFLEX correlation function as for Fig.~\ref{cdm1} but
compared with two open CDM models for which $\Omega_m=0.3$,
$\Omega_{\Lambda}=0.0$
(OCDM) and $\Omega_m=0.3$, $\Omega_{\Lambda}=0.7$ ($\Lambda$CDM), also
from Moscardini et al. (2000b).}
\label{cdm2}
\end{figure}
\vspace{0cm}

\begin{figure}
\vspace{-9.0cm}
\hbox{\hspace{0cm}\vspace{6.0cm}\psfig{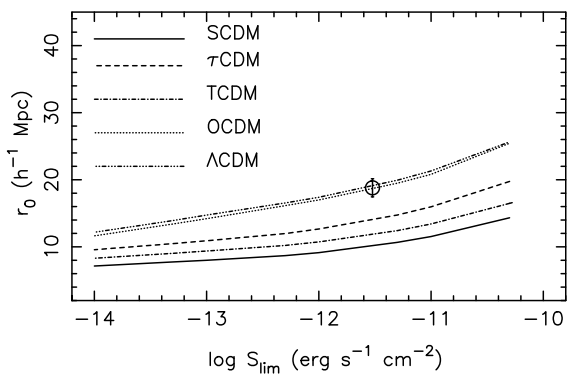}}
\caption{Comparison of the REFLEX correlation length $r_0$ from the
likelihood analysis with
predictions as a function of limiting X-ray flux for a range of CDM-type
cosmological models (see text for details). The REFLEX point is
plotted at
$3\times 10^{-12}$ erg s$^{-1}$ cm$^{-2}$. The likelihood error bar of
$0.9$ on $r_0$ has been
increased by a factor 1.5 to 1.35 here to account for the
extra contribution due to cosmic variance as described in Section 3.4.}
\label{cdmr0}
\end{figure}
\vspace{0cm}

\begin{figure}
\vspace{-9.0cm}
\hbox{\hspace{0cm}\vspace{6.0cm}\psfig{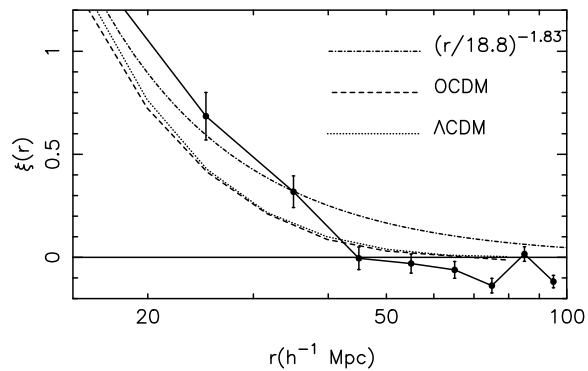}}
\caption{The REFLEX correlation function at low amplitude. The error
bars are calculated from eqn.~\ref{poisson}. Also shown is the best 
power-law fit over the 
range $4-40h^{-1}$ Mpc corresponding to $r_0=18.8$ and $\gamma=1.83$.
Predictions for the open and $\Lambda$-dominated CDM models from
Moscardini et al. (2000b) are also presented.}
\label{cross}
\end{figure}
\vspace{0cm}

Predictions for the clustering properties of X-ray selected clusters
from a
number of surveys, including REFLEX, have recently been made by
Moscardini et al. (2000b). In these predictions the
structures on a given scale are assumed to evolve by hierarchical
merging of smaller units and instantaneous merging on cluster scales. 
The comoving mass function of haloes is
computed using
the Press-Schechter (1974) technique but incorporating more recent
corrections which improve the comparison of Press-Schechter with
numerical
simulations (e.g. Sheth \& Tormen 1999). The link between X-ray
luminosity and mass of the hosting dark matter halo begins with
the empirical relation between gas temperature $T$ and X-ray
luminosity $L_{bol}$

\begin{equation}
 T=A L^{\beta}_{bol}(1+z)^{-\nu}, \label{tl}
  \end{equation}

  with $A=4.2$ \& $\beta=1/3$, which is a good approximation for
  clusters (e.g. David et al. 1993; White, Jones \& Forman 1997;
  Markevitch 1998). The parameter $\nu$ describing the evolution of the
  $T-L_{bol}$ relation is constrained by Moscardini et al. (2000b) using the
  X-ray cluster number counts over the range $5\times10^{-13} -
  3\times10^{-11}$ erg s$^{-1}$ cm$^{-2}$  ($0.5-2.0$ keV) taken from the RASS1
  Bright Sample  (De Grandi et al. 1999) and the fainter ROSAT Deep
Cluster Survey  Rosati et al. (1998). It is possible to convert
the temperature estimates from eqn.~\ref{tl} to halo
mass assuming a virial isothermal gas distribution and spherical
collapse (e.g. Eke, Cole \& Frenk 1996). Finally, in order to make
predictions for the correlation function of the REFLEX survey, 
Moscardini et al. (2000b) incorporate the actual sky coverage of the survey shown in
Fig.~\ref{effective} for the passband $0.1 - 2.4$ keV in an identical
manner
to the procedure used for calculating $\xi_{cc}(r)$ using the REFLEX
data described in Section 3.1 above.

The behaviour of the cluster correlation function for a range of
popular cosmological models based around cold dark matter (CDM) are
shown in
Fig.~\ref{cdm1} and Fig.~\ref{cdm2}.
The model predictions are taken directly from
Moscardini et al. (2000b) and represent: Standard (SCDM), $\tau$
($\tau$CDM) and tilted (TCDM) models, all with $\Omega_{m}=1$ and
$\Omega_{\Lambda}=0.0$; along with an open model $\Omega_{m}=0.3$
(OCDM) and a $\Lambda$-dominated model with $\Omega_{m}=0.3$ and
$\Omega_{\Lambda}=0.7$. An examination of these figures reveals clear
evidence for an inconsistency between all $\Omega_{m}=1$ models and the
REFLEX correlation function, with a significantly better fit to the
data for open or $\Lambda$-dominated cosmologies. This is further
illustrated in Fig.~\ref{cdmr0} which shows the Moscardini et al.
(2000b) predictions of correlation length with limiting X-ray flux.
While the Einstein-de Sitter models predict $r_0=11-13h^{-1}$Mpc, both
the OCDM and $\Lambda$CDM models predict $r_0\simeq20h^{-1}$Mpc. These
results are in agreement with the analysis of the X-ray clusters based
on the RASS1 Bright Sample (Moscardini et al. 2000a) and the
digitised optical surveys (e.g. Croft et al. 1997).

Confirmation of the general conclusions on the form
of the cosmological power spectrum comes from the behaviour of 
of $\xi_{cc}$ on large scales. In Fig. ~\ref{cross} we show the REFLEX
$\xi_{cc}(r)$ at low amplitude which shows a positive clustering 
signal out to at least $40h^{-1}$ Mpc ($\xi(30-40)=0.32\pm0.08$), with a  
zero-crossing $\simeq 45h^{-1}$
($\xi(40-50)=-4.7\times10^{-3}\pm0.05$). On larger scales the 
amplitude remains slightly negative ($\xi(50-100)=-0.07\pm0.02$). 
Also shown is the
curve representing the power law with $r_0=18.8h^{-1}$ Mpc and
$\gamma=1.83$, along with predictions from Moscardini et al. (2000b)
for the OCDM and $\Lambda$CDM models. Since the SCDM
model predicts a zero-crossing near $r\simeq33h^{-1}$Mpc (Klypin \& Rhee 1994),
the REFLEX data again support the findings of other 
cluster surveys that models with more power
than SCDM are required to adequately fit the large-scale
$\xi_{cc}(r)$. Generally for CDM-like models (with $n=1$ for the 
primordial spectral
index) the first zero-point occurs at $r\simeq16.5(\Omega_{m}h^2)^{-1}$
for a vanishing baryon fraction (see Klypin \& Rhee 1994). For the
particular parameterisation used by Moscardini et al. (2000b) the OCDM and 
$\Lambda$CDM models remain positive until $\simeq80h^{-1}$ Mpc, however
the small 
clustering amplitude in the data between $\simeq45-100h^{-1}$ Mpc seen in
Fig.~\ref{cross} prohibits any definitive comparison of the zero
crossings. 


%

\section{Summary}

Catalogues of galaxy clusters based on their X-ray emission 
provide a powerful tool for studies of large-scale structure. 
We present the spatial correlation function of the REFLEX cluster
survey, which consists of 449 X-ray emitting clusters above a flux
limit $3\times10^{44}$ erg s$^{-1}$ cm$^{-2}$ and covering a contiguous area
of 4.24 sr in the southern hemisphere. The advantages of X-ray
selection, combined with the increased statistics and 
high completeness of REFLEX enable a significant step to be taken in
establishing the clustering properties of clusters in the local universe. 
Over the scale $4-40h^{-1}$ Mpc we find a correlation
amplitude $r_0=18.8\pm0.9$ and power law index 
$\gamma=1.83^{+0.15}_{-0.08}$ for the entire survey. 
The high degree
of isotropy in the correlation
function demonstrates that systematic projection effects are not
present in the data. By analysing
volume-limited sub-samples we find no significant trend of  
clustering amplitude with X-ray luminosity. Comparing the REFLEX $\xi_{cc}$ 
results with predictions from
various CDM-type models which incorporate directly the areal coverage
of REFLEX, $\Omega_m\simeq0.3$ models provide an excellent fit, while
$\Omega_m=1$ \& $\Omega_{\Lambda}=0$ models fail to provide enough 
large-scale power. Finally, it is intriguing to note the concensus emerging 
between clustering studies and the lack of evolution in the 
abundance of X-ray clusters (e.g. Burke et al. 1997, Collins et al.
1997, Borgani et al. 1999, Henry
1997, Nichol et al. 1999), which also indicates that the 
Einstein de-Sitter universe is in trouble.

\section*{ACKNOWLEDGMENTS} 

We would like to thank the ROSAT team at MPE for providing the RASS
data ahead of publication and the COSMOS team at the ROYAL Observatory
Edinburgh for the digitised optical data. We are also indebted to
Rudolf D\"{u}mmler, Harald Ebeling, Alastair Edge, Andrew Fabian,
Herbert Gursky, Silvano Molendi, Marguerite Pierre, Waltraut Seitter, Giampaolo
Vettolani, and Gianni Zammorani for their help in the
observations taken at ESO and their work in the early stages of the
project. We also thank Kathy Romer for providing unpublished redshifts
and useful discussions. We thank Stefano Borgani for useful
feedback after reading a draft of this manuscript and also thank 
Lauro Moscardini for providing the CDM predictions in tabulation form. 
CAC acknowledges support from a PPARC Advanced
Fellowship during part of the lifetime of this project.

\section*{REFERENCES}

\myref{Abadi, M., Lambas, D., Muriel, H., 1998, ApJ., 507, 526}

\myref{Abell, G.O., 1958, ApJS, 3, 211}

\myref{Abell, G.O., Corwin, H.G., Olowin, R.P., 1989, ApJS, 70, 1}

\myref{Bahcall, N.A., 1988, ARA\&A, 26, 631}

\myref{Bahcall, N.A., Soneira, R.M., 1983, ApJ., 270, 20}

\myref{Bahcall, N.A., Soniera, R.M., Raymond, M., Burgett, W.S., 1986,
ApJ, 113, 15}

\myref{B\"{o}hringer, H., Guzzo, J., Collins, C.A., Neumann, D.M.,
Schindler, S., Schuecker, P., Cruddace, R.G., De Grandi, S.,
Chincarini, G., Edge, A.C., MacGillivray, H.T., Shaver, P., Vettolani,
G., Voges, W., 1998, The Messenger, No. 94, 21}

\myref{B\"{o}hringer, H., Voges, W., Huchra, J.P., McLean, B.,
Giacconi, R., Rosati, P., Burg, R., Mader, J., Schuecker, P., Simic,
D., Komossa, S., Reiprich, T.H., Retslaff, J., Tr\"{u}mper, J., 2000,
ApJ, submitted 2000a (astro-ph/0003219)} 

\myref{B\"{o}hringer, H., Schuecker, P., Guzzo, L., COllins, C.A.,
Voges, W., Schindler, S., Neumann, D.M., Chincarini, G., Cruddace,
R.G., De Grandi, S., Edge, A.C., MacGillivray, H.T., Shaver, P., 2000,
A\&A, submitted 2000b {\bf Paper I}}

\myref{Borgani, S., Plionis, M., Kolokotronis, V., 1999, MNRAS, 305,
866}

\myref{Borgani, S., Rosati, P., Tozzi, P., Norman, C., 1999, ApJ.,
517, 40}

\myref{Burke, D.J., Collins, C.A., Sharples, R.M., Romer, A.K., 
Holden, B.P., Nichol, R.C., 1997, ApJ, 488, L83}

\myref{Collins, C.A., Burke, D.J., Romer, A.K., Sharples, R.M.,
Nichol, R.C., 1997, ApJ, 479, L117}

\myref{Croft, R.A.C., Efstathiou, G., 1994, MNRAS, 267, 390}

\myref{Croft, R.A.C., Dalton, G.B., Efstathiou, G., Sutherland, W.J.,
Maddox, S.J., 1997, MNRAS, 291, 305} 

\myref{Dalton, G.B., Efstathiou, G., Maddox, S.J., Sutherland, W.J.,
1992, ApJ., 424, L1}

\myref{Dalton, G.B., Croft, R.A.C., Efstathiou, G., Sutherland, W.J.,
Maddox, S.J., Davis, M., 1994, MNRAS, 271, 47}

\myref{David, L.P., Slyz, A., Jones, C., Forman, W., Vrtilek, S.D.,
Arnaud, K.A., 1993, ApJ., 412, 479}

\myref{Davis, M., Peebles, P.J.E., 1983, ApJ, 267, 465}

\myref{De Grandi, S., Molendi, S., B\"{o}hringer, H., Chincarini, G.,
Voges, W., 1997, ApJ, 486, 738}

\myref{De Grandi, S., Guzzo, L., B\"{o}hringer, H., Molendi, S.,
Chincarini, G., Collins, C., Cruddace, R., Neumann, D., Schindler, S.,
Schuecker, P., Voges, W., 1999, ApJ, 513, 17}

\myref{Dekel, A., Bertschinger, E., \& Faber, S.M., 1990, ApJ., 364,
349}

\myref{Dickey, J.M., Lockman, F.J., 1990, ARAA, 28, 215}

\myref{Ebeling, H., Voges, W., B\"{o}hringer, H., Edge, A.C., Huchra,
J.P., Briel, U.G., 1996, MNRAS, 281, 799}

\myref{Ebeling, H., Edge, A.C., Fabian, A.C., Allen, S.W., Crawford,
C.S., B\"{o}hringer, H., 1997, MNRAS, 479, 101}

\myref{Efstathiou, G., Dalton, G.B., Sutherland, W.J., Maddox, S.J.,
1992, MNRAS, 257, 125}

\myref{Eke, V.R., Cole, S., Frenk, C.S., 1996, MNRAS, 282, 263}

\myref{Fisher, K.B., Davis, M., Strauss, M.A., Yahil, A., Huchra, J.,
1994, MNRAS, 266, 50}

\myref{Guzzo, L., B\"{o}hringer, Schuecker, P., Collins, C.A.,
Schindler, S., Neumann, D.M., De Grandi, S., Cruddace, R.G.,
Chincarini, G., Edge, A.C., Shaver, P., Voges, W., 1999, The
Messenger, No. 95, 27}

\myref{Guzzo, L., Bartlett, J.G., Cappi, A., Maurogordato, S., Zucca,
E., Zamorani, G., Balkowski, C., Blanchard, A., Cayatte, V.,
Chincarini, G., Collins, C.A., Maccagni, D., MacGillivray, H.,
Merighi, R., Mignoli, M., Proust, D., Ramella, M., Scaramella, R.,
Stripe, G.M., Vettolani, G., 2000, A\&A, 355, 1} 

\myref{Hamilton, A.J.S., 1993, ApJ., 406, L47}

\myref{Henry, J.P., 1997, ApJ, 489, L1}

\myref{Klypin, A., Kopylov, A.I., 1983, Sov. Astron. Lett., 9, 41}

\myref{Klypin, A., Rhee, G., 1994, ApJ, 428, 399}  

\myref{Lahav, O., Edge, A.C., Fabian, A.C., Putney, A., 1989, MNRAS,
238, 881} 

\myref{Ling, E.N., Frenk, C.S., Barrow, J.D., 1986, MNRAS, 223, L21}

\myref{Lucey, J.R., 1983, MNRAS, 204, 33}

\myref{Markevitch, M., 1998, ApJ., 504, 27}

\myref{Miller, C.J., Batuski, D.J., Slinglend, K.A., Hill, J.M., 1999,
ApJ., 523, 492}

\myref{Miller, C.J., Ledlow, M.J., Batuski, D.J., 2000, MNRAS,
submitted (astro-ph/9906423)}

\myref{Mo, H.J., Jing, Y.P., B\"{o}rner, G., 1992, ApJ., 392, 452}

\myref{Mo, H.J., Jing, Y.P., White, S.D.M., 1996, MNRAS, 282, 1096}

\myref{Moscardini, L., Matarrese, S., De Grandi, S., Lucchin, F.,
2000, MNRAS, in press (2000a)} 

\myref{Moscardini, L., Matarrese, S., Lucchin, Rosati, P., 2000,
MNRAS, in press(2000b)}

\myref{Nichol, R.C., Briel, O.G., Henry, P.J., 1994, ApJ., 267, 771}

\myref{Nichol, R.C., Collins, C.A., Guzzo, L., Lumsden, S.L., 1992,
MNRAS, 255, L21}

\myref{Nichol, R.C., Romer, A.K., Holden, B.P., Ulmer, M.P., Pildis,
R.A., Adami, C., Merrelli, A., Burke, D.J., Collins, C.A., 1999, ApJ,
521, 21}

\myref{Peacock, J.A., West, M.J., 1992, MNRAS, 259, 494}

\myref{Postman, M., Huchra, J.P., Geller, M.J., 1992, ApJ., 384, 404}

\myref{Press, W.H., Schechter, P., 1974, ApJ., 187, 425}

\myref{Romer, A.K., Collins, C.A., B\"{o}hringer, H., Cruddace, R.G.,
Ebeling, H., MacGillivray, H.T. \& Voges, W., 1994, Nature, 372, 75}

\myref{Rosati, P., Della Ceca, R., Norman, C., Giacconi, R., 1998,
ApJ., 492, L21}

\myref{Saunders, W., Rowan-Robinson, M., Lawrence, A., 1992, MNRAS,
258, 134}

\myref{Schuecker, P., B\"{o}hringer, H., Guzzo, L., Collins, C.A.,
Neumann, D., Schindler, S., Voges, W., Chincarini, G., Cruddace, R.,
De Grandi, S., Edge, A., M\"{u}ller, V., Reiprich, T.H., Retzlaff, J.,
Shaver, P., 2000, A\&A, submitted {\bf Paper III}}

\myref{Seth, R.K., Tormen, G., 1999, MNRAS, 308, 119}

\myref{Stark, A.A., Gammie, C.F., Wilson, R.W., Bally, J., Linke,
R.A., Heiles, C., Hurwitz, M., 1992, ApJs, 79, 77}

\myref{Sutherland, W.J., 1988, MNRAS, 234, 159}

\myref{Sutherland, W.J., Efstathiou, G.P., 1991, MNRAS, 248, 159}

\myref{Tr\"{u}mper, J., 1993, Science, 260, 1769}

\myref{Tucker, D.L., Oemler, A. Jr., Kirshner, R.P., Lin, H.,
Schectman, S.A., Landy, S.D., Schechter, P.L., Muller, V., Gottlober,
S., Einasto, J., 1997, MNRAS, 285, L5}

\myref{Voges, W., Aschenbach, B., Boller, T., Br\"{a}uninger, H.,
Briel, U., Burkert, W., Dennerl, K., Englhauser, K., Gruber, R.,
Haberl, F., Hasinger, G., K\"{u}rster, M., Pfeffermann, E., Pietsch,
W., Predehl, P., Rosso, C., Schmitt, J.H.M.M., Tr\"{u}mper, J.,
Zimmermann, H.U., 1999, A\&A, 349, 389}

\myref{White, D.A., Jones, C., Forman, W., 1997, MNRAS, 292, 419}
\end{document}